\documentclass{emulateapj}                                            %

\usepackage{epsfig}

\newcommand{\be}{\begin{equation}}
\newcommand{\ee}{\end{equation}}
\newcommand{\ba}{\begin{eqnarray}}
\newcommand{\ea}{\end{eqnarray}}

\newcommand{\saxj}{\mbox{SAX~J1808.4-3658}}

\newcommand{\ud}{\mathrm{d}}
\newcommand{\rb}{\bar{r}}
\newcommand{\textdegree}{\ensuremath{^\circ}}
\newcommand{\bigo}{\mathcal{O}}

\slugcomment{To appear in the Astrophysical Journal}
\shorttitle{Light Curves for Rapidly-Rotating Neutron Stars}

\begin{document}

\title{Light Curves for Rapidly-Rotating Neutron Stars}
\author{Coire Cadeau\altaffilmark{1}, Sharon M. Morsink\altaffilmark{1},
 Denis Leahy\altaffilmark{2}, and Sheldon S. Campbell\altaffilmark{1,3}}
\email{ccadeau@phys.ualberta.ca, morsink@phys.ualberta.ca, leahy@iras.ucalgary.ca, 
scampbel@physics.tamu.edu}

\altaffiltext{1}{Theoretical Physics Institute, Department of Physics,
University of Alberta, Edmonton, AB, T6G~2J1, Canada}
\altaffiltext{2}{Department of Physics and Astronomy, University of Calgary,
Calgary AB, T2N~1N4, Canada}
\altaffiltext{3}{Department of Physics, Texas A\&M University, College Station, TX, USA}

\begin{abstract}

We present raytracing computations for light emitted from the surface of a
rapidly-rotating neutron star in order to construct light curves for X-ray pulsars
and bursters. These calculations are for realistic models of rapidly-rotating neutron stars
which take into account both the correct exterior metric and the oblate shape of the star.
We find that the most important effect arising from rotation comes from the oblate shape
of the rotating star. We find that approximating a rotating neutron star as a sphere 
introduces serious errors in fitted values of the star's radius and mass if the 
rotation rate is very large. However, in most cases acceptable fits to the ratio  
$M/R$ can be obtained with the spherical approximation. 
\end{abstract}


\section{Introduction}
\label{s:intro}

One of the most fundamental problems in neutron star astrophysics is the determination
of the mass-radius relation through observations. This would allow the observational
determination of the equation of state (EOS) of cold supernuclear density material. While
the measurement of mass is possible if the star is in a binary, neutron stars are too small
to allow a direct measurement of their radii. One promising indirect method for inferring
the radius is through observations and modeling of light curves for X-rays emitted
from X-ray pulsars and neutron stars that exhibit type I X-ray bursts. This method requires
the raytracing of photons emitted from the star's surface in order to predict the signal detected by
the observer. In this paper we present the first raytracing calculations for 
rapidly-rotating neutron stars that include the correct metric and correct shapes of the star,
as well as allowing for arbitrary emission and detection directions. 

Rapidly-rotating neutron stars are a promising target for an effort to constrain the 
EOS since observations of their light curves could potentially allow the determination
of a star's mass and radius. In the case of a slowly rotating neutron star, the light curve
characteristics are controlled by the ratio $M/R$, but the value of $M$ or $R$ for the
star is not observable through the light curve. However, if the star is rapidly-rotating, the
Doppler boost resulting when the star's equatorial velocity is of the order $v/c \sim 0.1$ or
larger creates an asymmetry in the light curve which makes it possible to measure $v$. Since the
spin frequency is always known, the star's radius can be extracted. Knowledge of $R$ 
and $M/R$ could then potentially be used to constrain the EOS. The purpose of this paper is to 
determine how accurately these two parameters can be determined through light curve fitting.

In this paper we investigate the effect of rapid rotation on pulse shapes. In order to isolate
these effects, we have chosen to simplify our treatment of the shape, size, location and 
emissivity of the emitting region. It should be understood that the effects due to rapid 
rotation discussed in this paper are only one ingredient in the modeling of pulse shapes
and it will be necessary to combine these effects with more realistic treatments of the 
emitting regions in order to model real data. 

In order to proceed with a program of fitting light curves in order to infer the radius of a 
neutron star a number of assumptions must be made. The most fundamental assumption
made in this paper is that 
the light originates from the surface of the star and that no material lying between the star
and observer causes scattering. 
We make this assumption so that we can focus on the effects
due to rotation on the pulse shape, but it is possible that light is emitted from an 
extended region off of the star's surface and that the light interacts with the matter 
surrounding the star. For example, in the case of the slowly rotating X-ray pulsars 
complicated accretion columns are needed in order to fit the observations \citep{lea04}.
Pulse shape models of real data should allow for the possibility of extended emission 
regions.

Another important input to light curve models is the spectrum and emissivity of the emitting 
region. 
In the case of the 2.5 ms X-ray pulsar
\saxj, \cite{PG03} have shown that a hybrid spectrum including blackbody and Comptonized 
emission is needed. In their models,
thermal emission is directed towards the normal to the surface and
the Comptonized emission is directed away from the normal, creating a strong 
anisotropy.
Other models for the pulse shapes observed in type I X-ray bursts 
suggest  anisotropic emission as well (\cite{ML98}, \cite{Bhat05}).
Any anisotropy in the emission is known to have a strong effect on the 
light curve, so realistic light curve models must include the effects of 
anisotropy. 
In our present work, we have simplified our models to only 
allow isotropic emission. We expect that the magnitude of the rotational
effects discussed in this paper are similar to the magnitude of the effects 
caused by anisotropic emission. 

The assumed shape of the emitting region also strongly affects the pulse profile. 
Most studies have focused on either infinitesimal spots or simple Gaussian profiles. Recently
\cite{KR05} have produced
MHD simulations of accretion flows onto neutron stars that show the hot spots on
X-ray pulsars could have more complicated shapes, confirming predictions
made by \cite{GL78}. For the case of type I X-ray bursts
it is possible for normal modes to be excited \citep{Heyl04}, which would produce patterns 
associated with the modes \citep{Heyl05,LS05}. In our calculations we have chosen the 
simplest possible pattern (an infinitesimal uniform brightness spot) in order to focus 
on the effects due to rotation. Models of real data must allow for more complicated
emission shapes and brightness patterns. 

The gravitational field outside of the star is important, for it gravitationally redshifts
the photons and bends their direction of propagation. For non-rotating stars the 
Schwarzschild metric is all that is required to describe the photon trajectories 
\citep{PFC83}. The Schwarzschild + Doppler (S+D) approximation \citep{ML98,PG03}
was introduced as a simple method for calculating the light curves of slowly 
rotating stars. In the S+D approximation
the gravitational field is modeled by the static Schwarzschild solution and the
rotational effects are approximated by using the special relativistic Doppler transformations.
One of the goals of this paper is to find the conditions under which the S+D approximation
will fail. For rapidly-rotating neutron stars, the metric must be computed numerically
using one of many known algorithms, such as that of \cite{CST94}. 
The Kerr black hole metric can be made to agree to first order in 
angular velocity with the metric of a rotating neutron star metric by choosing
the Kerr parameter $a$ to be $J/M$ where $J$ is neutron star's angular momentum. 
The Kerr black hole is a poor approximation to a rotating 
neutron star for spin frequencies above 400 Hz (for an example involving the innermost
stable circular orbit, see Miller et al 1998). However, some calculations 
(Braje et al 2000) have suggested that the Kerr metric may suffice for 
raytracing applications. This motivates \citep{BRR00,Bhat05} the use of the 
Spherical Kerr approximation (SK) where a spherical
surface is embedded in the Kerr spacetime. In this paper we will show that when the 
S+D approximation fails, the SK approximation also fails, so that the SK approximation
is not necessarily a better choice than S+D. In this paper we compute the metric of
rapidly-rotating neutron stars using realistic equations of state and perform raytracing
on the computed numerical spacetime. In a previous paper \citep{CLM05} (CLM) we performed
similar calculations that were restricted to the equatorial plane. In this paper we
generalize to allow arbitrary initial spot locations and arbitrary observer locations.

The final important input to a raytracing program is the shape of the star's surface,
which for a rotating fluid star is an oblate spheroid.
In almost all previous calculations it has been assumed that the surface of the star is
a sphere, irrespective if the exterior gravitational field was assumed to be Schwarzschild
or Kerr. We are aware of only one previous calculation, by \citet{BR01} that takes the
oblate surface into account. In that paper they consider further scatterings of photons
within the magnetosphere, so that the effect of oblateness is not clearly separated 
from other effects. In order to motivate the choice of a spherical surface, it is often stated that
the ratio of polar axis to equatorial axis is very close to unity, even for rapid rotation.
However, this ratio is neither coordinate invariant, nor directly observable. The 
oblate shape of the star has two main effects. (I) The gravitational field near the 
the poles is stronger than near the equator, so the redshift and photon deflection
are larger for photons emitted near the poles. (II) Light is emitted at different 
angles with respect to the normal to the surface if the surface is an oblate spheroid
instead of a sphere.  As a result some parts of a spherical star that an observer can't see will be
visible if the star is actually oblate (the converse is also true). Of these 
effects, the second is the most important, and leads to failure of the S+D and SK
approximations for rapid rotation.  

Another application which makes use of raytracing is the calculation of absorption line
profiles. Absorption line profiles have been computed in the past using the S+D 
approximation \citep{OP03,VS04} and using the SK approximation \citep{BML06}. 
Interestingly, \citet{BML06} have identified signs of frame-dragging in the line profiles
computed using the Kerr metric for very rapidly-rotating neutron stars. \citet{CMBW06}
compared line profiles using oblate stars (using the same type of method used in this
paper) with the line profiles calculated in the spherical approximation and found that 
the differences are insignificant for line profile calculations. However, it should be noted
that the calculations done by \citet{CMBW06} assumed that light from all of the star 
contributes to the line profile, which would tend to average out differences arising
from the shape of the star.

In this paper we wish to explore the rapid rotation effects of the metric and oblateness
on the calculated light curves of neutron stars and on the fitting of a star's mass and 
radius from the light curves. In order to isolate these effects, we consider the simplest
possible assumptions for all other aspects. We assume isotropic emission from an
infinitesimal spot, but allow for spots and observers at any location. 
As discussed in the previous paragraphs, there is enormous uncertainty in
the shape, size and emissivity of the emitting regions which will have the 
effect of making it difficult to discern the relativistic effects discussed 
in this paper. However, we will show that for very rapidly rotating neutron
stars, the relativistic rotational effects are large and should be incorporated 
in models of real data.
In section
\ref{s:light} we describe the method used to compute the light curves for rotating 
neutron stars. In section \ref{s:approx} we outline the four types of approximation
schemes that could be used to produce light curves. In section \ref{s:results} we
compare the light curves from the exact and approximate methods. We show how
programs that attempt to fit light curves using either spherical approximation 
will converge to incorrect values of mass and radius at spin frequencies larger than about
300 Hz, although they may converge to the correct value of the ratio $M/R$. 
We conclude with a discussion of these results in section \ref{s:conclusions}.

\section{Light Curve Calculations for Rapid Rotation}
\label{s:light}

The metric of a rotating neutron star can be accurately 
computed using a public-domain code 
{\tt rns}\footnotemark 
\footnotetext{Code available at http://www.gravity.phys.uwm.edu/rns/}
 \citep{SF95} based on the code by \cite{CST94}. This code computes the
four metric potentials $\alpha, \gamma, \rho$ and $\omega$, that appear in 
the general stationary axisymmetric metric 
\be 
ds^2
      = -e^{\gamma+\rho}dt^2+e^{2\alpha}\left(d{\rb}^2+{\rb}^2d\theta^2\right)+
e^{\gamma-\rho}{\rb}^2\sin^2\theta(d\phi-\omega dt)^2,
\label{metric}
\ee
where the metric potentials are functions of only ${\rb}$ and $\theta$.
The interpretation of these potentials and coordinates has been discussed in more
detail in CLM.

We have chosen two equations of state (EOS) from the \cite{AB77} catalog which 
span a realistic range of stiffness (although we have not included quark stars).  
EOS A is one of the softest EOS and 
EOS L is one of the stiffest. While these EOS are rather old-fashioned and the
physical assumptions made in deriving these EOS are now considered too simplistic,
these EOS do provide a wide span of properties that allow us to illustrate the 
different levels that rotation can affect the resulting light curves. Modern 
EOS, such as EOS APR (\cite{APR98}) have properties that are bracketed by the older
EOS A and L.

We have restricted our computations to stars with masses
of $1.4 M_\odot$ and spin frequencies from 100 to 600 Hz in 100 Hz increments
in order to explore the spin dependent effects. The 1.4 $M_\odot$ stars constructed 
from EOS A have equatorial radii ranging from 9.5 to 9.8 km (depending on the spin rate)
allowing for compact stars. The 1.4 $M_\odot$ stars constructed with EOS L have 
equatorial radii ranging from 14.8 to 16.4 km allowing for larger stars that are 
strongly affected by rotation. In contrast to the extreme EOS used in our calculations,
a $1.4 M_\odot$ neutron star constructed from the modern EOS APR has an 
equatorial radius ranging from 11.4 to 11.8 km as the spin is increased from 0 Hz to 
600 Hz. The properties of the neutron star models used in this
paper are summarized in Table 1. In Table 1 we show the values of the compactness 
$M/R$ and the speed of the star $v$ at the equator. The stars constructed with EOS A
are more compact than the EOS L stars, so the effects of light-bending and time-delays
are most important in the EOS A stars. The equatorial velocities are highest for the
the larger EOS L stars. As an alternative to the equatorial velocity of the star, the
dimensionless rotation parameter $j=J/M^2$ (where $J$ is the star's angular momentum) 
is a useful measure of the importance of
rotation for the star. The effect of oblateness is quadratic in the parameter $j$, so
the EOS L stars with high spin frequencies are very oblate while the more compact 
stars are only slightly oblate. In addition, if the neutron star has a mass larger than
$1.4 M_\odot$ then it will be more compact and less affected by rotation. 

The fastest neutron stars in X-ray binaries have spin frequencies near 600 Hz, motivating 
our range of spin frequencies.
The recent discovery \citep{Hessels} of a 716 Hz binary radio pulsar suggests that 700 Hz X-ray pulsars 
may be discovered in the future, but for the present purposes, stars
spinning at 600 Hz serve to illustrate the largest effects due to rotation.

\subsection{Geodesic Equations}

The geodesic equations for the coordinate positions of a photon depend
on the impact parameter $b$ defined by the ratio of the photon's angular momentum
to its energy. The equations are integrated with respect to an
affine parameter $\lambda$ scaled so that photon propagation is
independent of the photon's energy. We use an overdot to denote differentiation
with respect to $\lambda$. The form of the geodesic equations used in 
our integration is
\begin{eqnarray}
  \dot{t}&=&e^{-(\gamma+\rho)}(1-\omega b)\label{teq}\\
  \dot{\phi}&=&\omega e^{-(\gamma+\rho)}(1-\omega b)+e^{\rho-\gamma}\frac{b}{\rb^2\sin^2\theta}\label{phieq}\\
  \ddot{\rb} &=&  - \alpha_{,\rb}\left( \dot{\rb}^2 -\rb^2\dot{\theta}^2\right)
  -2 \alpha_{,\theta} \dot{\rb}\dot{\theta} + \rb \dot{\theta}^2
+      \frac12 e^{-2\alpha} \mathcal{B}_{,\rb}
     \label{rdoteq}\\
  \ddot{\theta}&=& \alpha_{,\theta}\left( \frac{\dot{\rb}^2}{\rb^2} - \dot{\theta}^2\right)
                - 2 \left(\alpha_{,\rb}+\frac{1}{\rb}\right)\dot{\rb}\dot{\theta} 
       	+ \frac{1}{2\rb^2} e^{-2\alpha} \mathcal{B}_{,\theta} 
         \label{mudoteq}\\
  \mathcal{B} &=& 
e^{-(\gamma+\rho)}(1-\omega b)^2  - \frac{b^2 e^{\rho-\gamma}}{\rb^2 \sin^2\theta}           
\label{Beq}.
\end{eqnarray}
These equations are overspecified, since the fact that photon's 
four-velocity vector has vanishing norm leads to the  momentum constraint 
\begin{eqnarray}
  \dot{\rb}^2+{\rb^2}\dot{\theta}^2&=&e^{-2\alpha} {\mathcal B}(\rb,\theta) \equiv {\mathcal A}(\rb,\theta)  
\label{eq:nullconst}
\end{eqnarray}
which provides an extra relation between $\dot{\rb}$ and $\dot{\theta}$, 
and defines the function $\mathcal{A} = e^{-2\alpha} \mathcal{B}$. While it would be possible
to eliminate equation (\ref{mudoteq}) using equation (\ref{eq:nullconst}), this would not
allow us to evaluate the accuracy of the code. We have chosen to integrate equations
(\ref{teq}) - (\ref{mudoteq}) and use the momentum constraint equation as a test
of the code's accuracy. Our code uses a Runge-Kutta algorithm with
adaptive step size.

The main addition to the existing
{\tt rns} code which we made was a calculation of the derivatives of the metric potentials which are
required for the geodesic integration. A simple finite differencing scheme will fail at the poles of
the star and also tends to unacceptably large errors far from the star. Instead, we take advantage of
the explicit sum formulae \citep{KEH89} for the potentials and explicitly take their first derivatives with 
respect to $\bar{r}$ and $\theta$ as well as the mixed second derivative. This provides a smooth set
of potentials and derivatives at discrete grid points which can be used in a bicubic interpolation scheme
\citep{Press}
to find the values of the potentials and their derivatives at intermediate points. 

\subsection{Initial Conditions}
\label{ss:ic}

The surface of the star is described by the function $\rb_s(\theta)$, which can be
found numerically. Given an initial value of co-latitude $\theta = \theta_i$, the
initial value of the radial coordinate is $\rb_i = \rb_s(\theta_i)$ for photons
emitted from the surface of the star. With these initial values of the coordinates,
the positivity of the right-hand side of
equation~(\ref{eq:nullconst}) yields a constraint on the allowed values
of $b$, $b_- \leq b \leq b_+$, with
\be
\label{eq:bconstr}
b_\pm = \pm \frac{ e^{-\rho} \rb_i \sin\theta_i }{ 1 \pm \omega e^{-\rho}\rb_i\sin\theta_i },
\ee
where the metric potentials are to be evaluated at the initial
coordinate.

\begin{figure}
\begin{center}
\plotone{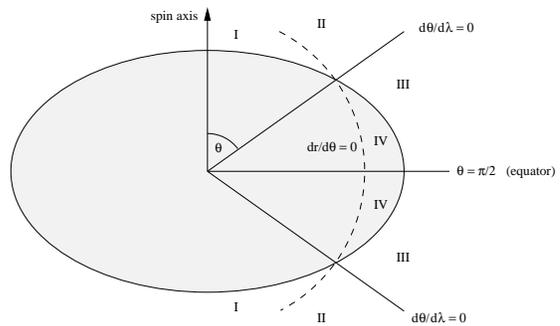}
\end{center}
\caption[]{ Side view of a rotating neutron star (solid curve). 
Photons emitted from a point on the star at an angle $\theta$ from
the spin axis can be emitted into regions 
I, II, or III if the star's surface is an oblate spheroid. If the star's surface is
spherical (shown as a dashed curve), photons can only be emitted into regions II, III, or IV.
}
\label{fig:ics}
\end{figure}

With an initial point and a value of $b$, we
can carry on to calculate the allowed values of
$\dot{\theta}_i$. Rewriting equation~(\ref{eq:nullconst}), we have that
\be
\label{eq:nullcond2}
\dot{\theta}^2 \left[ \left(\frac{ \ud \rb }{ \ud \theta }\right)^2 + \rb^2\right] =
 \mathcal{A}(\rb,\theta),
\ee
where $\dot{\rb}/\dot{\theta} = \ud \rb/\ud \theta$. The allowed values of 
$\dot{\theta}_i$ follow from an analysis of the extreme values of the
term in parentheses on the left-hand side of
equation~(\ref{eq:nullcond2}). For light rays emitted tangent to the
star's surface, $d\rb/d\theta = d\rb_s/d\theta \equiv \rb'_s(\theta)$.
For a spherical surface one expects $\dot{\rb} \geq 0$
for outgoing rays, but since we are considering stars that are
(perhaps very slightly) oblate, there are 
``glancing'' rays with $\dot{\rb}_i < 0$.  Figure~\ref{fig:ics} shows
the situation for points above and below the equatorial plane in three
separate regions where outgoing rays can be defined. 
In each region,
the sign of $\pm$ is chosen to match the sign of
$\cos\theta_i$. Evaluating all quantities at the initial
point, we have the following situations in Figure~\ref{fig:ics}:
\begin{description}
\item[Region I.] Rays with $\dot{\rb}_i < 0$ and
$\pm \dot{\theta}_i < 0$. In this region we have 
\be
\frac{ \mathcal A (\rb_i,\theta_i) }{ \rb_i^2 + \left(\rb'_{s}(\theta_i)\right)^2 }
\leq \dot{\theta}_{i}^2 
\leq
\frac{ \mathcal A (\rb_i,\theta_i)} { \rb_i^2 }.
\ee
This region contains the rays that would be prohibited if
the surface of the star were a sphere with radius $\rb = \rb_i$.
\item[Region II.] Rays with $\dot{\rb}_i > 0$ and $\pm \dot{\theta}_i
< 0$. In this region,
\be
0
\leq 
\dot{\theta}_{i}^2 
\leq
\frac{ {\mathcal A} (\rb_i,\theta_i)} { \rb_i^2 }.
\ee
\item[Region III.] Rays with $\dot{\rb}_i > 0$ and $\pm \dot{\theta}_i
> 0$. In this region,
\be
0
\leq 
\dot{\theta}_{i}^2 
\leq
\frac{ \mathcal A  (\rb_i,\theta_i)} { \rb_i^2 + \left(\rb'_{s}(\theta_i)\right)^2 }.  
\ee
\item[Region IV.] Rays with $\dot{\rb}_i > 0$ and $\pm \dot{\theta}_i
> 0$. In this region,
\be
\frac{ \mathcal A  (\rb_i,\theta_i)} { \rb_i^2 + \left(\rb'_{s}(\theta_i)\right)^2 }
\leq 
\dot{\theta}_{i}^2 
\leq
\frac{ \mathcal A  (\rb_i,\theta_i)} { \rb_i^2 }.  
\ee
This rays in this region would only be allowed if the
surface of the star were a sphere with radius $\rb = \rb_i$.
\end{description}
Given an initial point on the star, only photons emitted into regions I, II, or III are
allowed if the surface is oblate. If a spherical approximation to the surface is being made, only
photons emitted into regions II, III, or IV are allowed.
Given a value of $\dot{\theta}_i$, the
corresponding value of $\dot{\rb}_i$ is fixed by equation~(\ref{eq:nullconst});
if necessary, the sign of $\dot{\rb}_i$ is disambiguated according to
which region in Figure~\ref{fig:ics} is considered.

Integration of a single null ray proceeds by setting the initial
coordinates $\rb_{i}$ and $\theta_i$, selecting an allowed value
of $b$, selecting an allowed value $\dot{\theta}_i$ according 
to the above prescription of the geometric constraints, and
fixing $\dot{\rb}_i^2$ and the sign of $\dot{\rb}_i$ by the 
momentum and geometric constraints. The differential
equations~(\ref{teq}), (\ref{phieq}), (\ref{rdoteq}),
and (\ref{mudoteq}) are then integrated numerically until the radial
coordinate reaches a predetermined large value.

\subsection{Redshift and Zenith Angle}

In CLM we presented equations for the photon's redshift and initial zenith angle as measured
in the frame co-rotating with the fluid for the case of photons emitted from and into the equatorial plane.
The equations for arbitrary initial conditions are derived in a similar manner.  The star's fluid has four-velocity
$u^{a}_{\star} = (t^{a}+\Omega_{\star}\phi^{a})/N_{\star}$ where the normalization function has the value
\be
N_{\star}^{2} = e^{\gamma+\rho}\left( 1 - v_{Z}^{2}\right).
\ee
where 
\be
v_Z = (\Omega_\star - \omega) e^{-\rho} \rb \sin\theta
\ee
is the speed of
the star's fluid as measured by a zero angular momentum observer.

Using $\ell^{a}$ to denote the photon's four-velocity (as defined in section 2.1), the photon's redshift
as measured by an observer at infinity is given by $1+z = - u_{\star}\cdot\ell$, where we use a ``dot'' to 
denote the inner product with respect to the full 3+1 spacetime metric given in equation (\ref{metric}).
The photon 
redshift is then
\be
1+z = e^{-\frac12(\gamma +\rho)} \frac{ (1 - \Omega_\star b)}
{
\sqrt{ 1 - v_Z^2 }
}. 
\label{eq:red}
\ee
where all quantities are evaluated at the initial location on the star,

An inertial observer at infinity measures time using the $t$ coordinate, so 
the star's spin period is
$2\pi/\Omega_\star$. An inertial observer sitting at the star's surface measures 
time with the local proper time $\tau$, where the proper and coordinate times
are related by
\be
\frac{d \tau}{dt} = e^{\frac12({\gamma+\rho})} \sqrt{1-v_Z^2}.
\label{dtaudt}
\ee
The observer at the star's surface 
measures the spin period to be gravitationally blue-shifted by the factor $d\tau/dt$.

Given a point on the surface of the star, the normal vector $n^a$ to the surface is 
defined by
\begin{eqnarray}
\label{eq:geomnormal:rb}n^{\rb} & = & 1 \\
\label{eq:geomnormal:th}n^{\theta} & = & - \frac{\rb_s'(\theta)}{\rb_s^2} .
\end{eqnarray} 
The angle between the initial photon direction and this normal, as measured in the
star's corotating frame is given by
\be
\cos \alpha_{e} = \frac{ n^{a} \ell^{b} h_{\star ab}} { n_{\star}\ell_{\star}}
\ee
where $h^{ab}_{\star} = g^{ab} + u^{a}_{\star}u^{b}_{\star}$ gives the projection of
four-vectors to the 3-space defined for an observer rotating with the star, 
and $n_{\star}=h_{\star ab} n^{a} n^{b}$ and $\ell_{\star}=h_{\star ab}\ell^{a}\ell^{b}$.
With these definitions, the angle $\alpha_{e}$ is
\be 
\cos \alpha_e = \frac{e^\alpha}{(1+z)}
{\left(\dot{\rb} - \dot{\theta} {\rb_s'}(\theta)\right)}
{\left(1 + \left(\frac{\rb_s'(\theta)}{\rb_s}   \right)^2 \right)^{-1/2}}.
\ee
In this equation, $\dot{\rb}$ and $\dot{\theta}$ are the initial values of the photon's
velocity components subject to the constraints given in section \ref{ss:ic}.

\subsection{Doppler Factors}

We expect that in the limit of slow rotation that the full formalism presented in this
paper should reduce to the S+D approximation. In order to show that this is the case,
we require the calculation of the special relativistic factors that enter into the Doppler
shift formula. The Lorentz ``boost'' factor $\eta$ is defined by
\be
\eta = \frac{\sqrt{1-v^2}}{1-v\cos \xi}
\ee
where $v$ is the speed of the emitting area, as measured by an inertial observer at the star,
and $\xi$ is the initial angle between the fluid velocity and the initial photon direction,
as measured by the inertial observer.

The inertial observer has four-velocity $u^a_i = t^{a}/N_{i}$, where the normalization factor is 
given by $N_{i} = e^{\frac12(\gamma+\rho)} \left( 1 - \omega^{2} e^{-2\rho}\rb^{2}\sin^{2}\theta\right)^{1/2}$.
The spatial projection operator for this observer is $h_{i}^{ab} = g^{ab}+u^{a}_{i}u^{b}_{i}$, allowing the
speed of the emitting area to be defined through the relation
\begin{eqnarray}
v &=& \frac{(h_{i ab} u^{a}_{\star} u^{b}_{\star} )^{1/2}}{| u_{i}\cdot u_{\star}|}\nonumber \\
&=& \Omega_{\star} e^{-\rho} \rb \sin\theta (1 - \omega(\Omega_{\star}-\omega)e^{-2\rho}\rb^{2}\sin^{2}\theta)^{1/2}
\end{eqnarray}
The angle $\xi$ is defined by
\be
\cos \xi = \frac{h_{i ab} u^{a}_{\star} \ell^{b}}
{ (h_{i ab}u^{a}_{\star}u^{b}_{\star})^{1/2} (h_{i ab}\ell^{a} \ell^{b})^{1/2} },
\ee
which leads to the expression
\be
1 - v\cos\xi = \frac{ \left(1 - e^{-2\rho}\rb^{2}\sin^{2}\theta \omega^{2} \right)}
{\left(1 - e^{-2\rho}\rb^{2}\sin^{2}\theta \omega(\omega-\Omega_{\star})\right)} (1-\Omega_{\star}b). 
\ee

We wish to make contact with the S+D approximation, in which the exterior gravitational field of the
star is approximated by the Schwarzschild metric.  The limiting values of the metric potentials for zero rotation
were given by equations (3) - (5) of CLM. As discussed in CLM, the S+D approximation can be 
found by taking the limit of the frame-dragging potential $\omega$ to zero and using the Schwarzschild limits for
the potentials $\gamma, \rho$ and $\alpha$. We will denote the S+D approximation 
of any expression with the limit $\omega \rightarrow 0$. In this limit, the values of $v$ and $v_{Z}$ are
\be
\lim_{\omega\rightarrow0} v = \lim _{\omega\rightarrow0} v_{Z} = \frac{\Omega_{\star}R \sin\theta}{\sqrt{1 -2M/R}},
\ee
and the Doppler Boost factor is
\be
\lim_{\omega\rightarrow0}\eta =  \frac{\sqrt{1-v^{2}}}{ 1 -\Omega_{\star} b}.
\label{eq:etaS+D}
\ee

>From equation (\ref{eq:red}) it can be seen that the redshift factor in the S+D approximation reduces to the 
expected expression
\be
\lim_{\omega\rightarrow0} (1+z) =  (1 - \frac{2M}{R})^{-1/2} \eta^{-1}.
\label{eq:zS+D}
\ee

\subsection{Time of Arrival and Azimuthal Deflection Angle}

The coordinate time of arrival for each photon is denoted $T$ and can be found by
integrating equation (\ref{teq}) and subtracting off the time of arrival for
some arbitrarily chosen reference photon. Similarly, the azimuthal deflection
angle $\psi$ (the change in $\phi$ coordinate) can be found by integrating
equation (\ref{phieq}). Both the time of arrival and the azimuthal deflection
angle can be written as functions of the photon's initial latitude on the star $\theta_{i}$,
the final latitude of the photon $\theta_{f}$ and the photon's impact parameter $b$.

Consider the family of photons all emitted from the same latitude on the star,
and all received by the same observer at infinity. These 
photons all have different values of $b$, and their times of arrival will be written
as $T(\theta_i, \theta_f, b)$ and their azimuthal deflection is $\psi(\theta_i,\theta_f,b)$. 
If $\theta_i$ and $\theta_f$ are fixed, two photons with impact parameters $b$ 
and $b+\Delta b$ have  times of arrival differing by
\be
\Delta T = \frac{d T}{d b} \Delta b.
\ee
The deflection angles for the two photons differ by
\be
\Delta \psi = \frac{d \psi}{d b} \Delta b.
\ee
In the Appendix, we show that $d T/d b = b d \psi/d b$. This
leads to the identity, for photons with identical initial and final latitudes
\be
\Delta T(\theta_i,\theta_f, b) = b \Delta \psi(\theta_i,\theta_f,b).
\ee

\begin{figure}
\begin{center}
\includegraphics[width=4in]{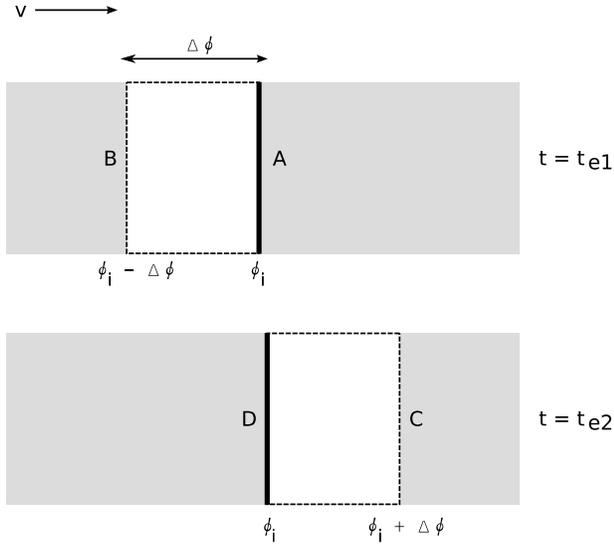}
\end{center}
\caption[]{An emitting region with azimuthal width $\Delta \phi$ which moves across the star
at constant latitude. Photons from locations A and B are emitted simultaneously at time $t_{e1}$ 
and later, at time $t_{e2}$, photons C and D are emitted.}
\label{fig:time}
\end{figure}

Consider two points on the same latitude of the star separated by an azimuthal
angle $\Delta \phi$ as shown in Figure~\ref{fig:time}. As the star rotates, the azimuthal 
locations of the two points 
change, but the separation remains fixed. Suppose that photons are emitted simultaneously
from the two points so that the leading photon (labeled ``A'') has impact parameter $b$ and is emitted at some
initial azimuthal angle $\phi_i$, and the trailing photon (labeled ``B'') has impact parameter $b + \Delta b$
and initial azimuthal angle $\phi_i - \Delta \phi$. (In our sign convention the impact parameter is 
positive on the blue side of the star, and the azimuthal coordinate $\phi$ is
negative on the blue side. 
With this sign convention $\Delta \phi$ is always positive.)
Define the time of arrival for the leading photon to be $T_A$. Then the time of arrival
for the trailing photon is $T_B = T_A + b \Delta \phi$ since  $\Delta \phi = - \Delta \psi$.
With our sign convention, on the blue side of the star $T_B>T_A$,
so that the leading photon arrives before the trailing photon. On the red side
of the star the opposite is true. 

The location of the two emitting points at a coordinate time $\Delta t_e$ later is shown in the second panel
of Figure~\ref{fig:time}. In this time interval each point has moved through an angle $\Delta \phi$, so the
lapse in coordinate time is
$
\Delta t_e = {\Delta \phi}/{\Omega_\star},
$
where the subscript $e$ is used to denote the lapse in coordinate time measured at the location of the
emitting points.
The two photons emitted simultaneously at this later time are now labeled ``C'' (for the leading photon) 
and ``D'' (for the trailing photon). The trailing photon arrives at time
\be
T_D = \frac{\Delta \phi}{\Omega_\star} + T_A
\ee
and the leading photon arrives at time
\be
T_C = T_D - b \Delta \phi = T_A + \frac{\Delta \phi}{\Omega_\star}\left(1 -b \Omega_\star\right).
\ee
The observer sees the photons move through the angle $\Delta \phi$ in the time interval $\Delta T$,
\be
\Delta T = T_C - T_A = \frac{\Delta \phi}{\Omega_\star}\left(1 -b \Omega_\star\right)
= \Delta t_e \left(1 -b \Omega_\star\right).
\label{deltaT}
\ee

\subsection{Solid Angle}

The flux detected from an emitting area on the star is directly proportional to the solid
angle subtended by the area, as measured by the observer. Consider any three photons arriving
at the detector. Since the observer is located in a flat region of space, the solid angle can
be found by first calculating the three angles separating the three photons. These three angles
form a triangle whose area can be computed using Euclidean geometry. 

Consider two photons seen by the observer with four-momenta $\ell$ and $m$ which
has components $\ell_t=m_t=-1$,  $\ell_\phi=b$,  $m_\phi = b+db$, $m^\theta=\ell^\theta + d\ell^\theta$
and $m^{\rb} = \ell^{\rb} + d\ell^{\rb}$. The angle between these two photons, as measured by 
a static observer at infinity can be found using a method similar to the Appendix of CLM,
except that we now allow for photons detected and emitted off of the equator. After some 
algebra, the angle $\epsilon$ is given by 
\be
\cos (\epsilon) 
= 1 + g_{tt} \left[ g^{\phi \phi} (\ud b)^2 + g_{\rb \rb} (\ud
  \ell^{\rb})^2 + g_{\theta \theta} (\ud \ell^\theta)^2  \right].
\ee
For small angular separations, $\cos(\epsilon) = 1 - \epsilon^2/2 + \bigo(\epsilon^4)$,
so the angle separating the photons is
\be
\epsilon = \sqrt{-g_{tt}} \left[ g^{\phi \phi} (\ud b)^2 + g_{\rb \rb} (\ud
  \ell^{\rb})^2 + g_{\theta \theta} (\ud \ell^\theta)^2  \right]^{1/2}.
\ee

Given the three angles $\epsilon_i$ (where $i$ runs from 1 to 3) separating the three
photons, the solid angle subtended by the photons can be found (assuming flat space
at infinity) from a Euclidean formula such as 
\be
d\Omega = \sqrt{ s (s-\epsilon_1) (s-\epsilon_2) (s-\epsilon_3) },
\ee
where $s=(\epsilon_1+\epsilon_2+\epsilon_3)/2$. 

In our calculations, we must be careful about whether the photons were emitted
or detected simultaneously. We use $d \Omega_e$ to denote the solid angle subtended
by photons that were emitted simultaneously, but detected at different times. 
The solid angle subtended by photons that are detected simultaneously but
emitted at different times is denoted $d \Omega_o$.

\subsection{Light Curves}
\label{s:lightcurves}

In order to construct a light curve, the flux as a function of observed coordinate
time must be calculated. Consider two different instantaneous definitions of the
flux from a small spot on the star with width $\Delta \phi$.
The specific flux detected at time $T$ is
\be
F_{o}(T) = \frac{I_{\nu_e}(\alpha_e)}{(1+z)^3} d\Omega_o,
\ee
where the values of $z$ and $\alpha_e$ have been chosen so that the photons arrive at the
observer at the correct time.
The flux received in the detector over a time interval due to photons emitted 
simultaneously at the same value of coordinate time $t_e$ is
\be
F_{e}(t_e) =  \frac{I_{\nu_e}(\alpha_e)}{(1+z)^3} d\Omega_e.
\ee

However, neither definition of flux describes what is truly measured in
the detector, since the detector measures flux over a finite time interval
$\Delta T$. Light detected during the interval $\Delta T$ is emitted over
the coordinate time interval $\Delta t_e$, so the total energy per unit area
that arrives in the detector is
\be
\hbox{Total Energy} = F_e(t_e) \Delta t_e ,
\ee
where we are now integrating the specific flux over all energies.
The flux in the detector is averaged over the collection time interval $\Delta T$,
so the measured flux is
\be
F(T,\Delta T) = \frac{\hbox{Total Energy}}{\Delta T} = 
F_e(t_e) \frac{\Delta t_e}{\Delta T}.
\ee
Making using of equation (\ref{deltaT}), the expression for the flux reduces to 
$F(T,\Delta T) = F_e(t_e)/(1-b\Omega_\star)$. 
In the S+D approximation (see equation (\ref{eq:etaS+D})) this yields
\be
\lim_{\omega\rightarrow0} F(T,\Delta T) = F_e(t_e) \eta, 
\ee
as derived using other methods by \citet{PG03}.

In CLM we erroneously omitted the extra boost factor derived in this section. As a result,
the pulse shapes shown in that paper are not correct. This means that in CLM we underestimated
the errors introduced by neglecting time delays into the S+D fitting program.
 However, the conclusions made by
CLM are unchanged: for equatorial photons, the most important effect is the binning of
photons by the correct arrival time.

\subsection{Approximation Schemes}
\label{s:approx}

The raytracing procedure discussed in this section is time-consuming, so simpler 
approximations based on analytic spacetime metrics are desirable. Two approximations
are commonly used, the Schwarzschild + Doppler (S+D), and the Spherical Kerr (SK) approximations.
In both of these approximations a black hole metric with a mass identical to the neutron
star's mass is used instead of the neutron star metric. In the case of the SK approximation,
a Kerr metric with angular momentum identical to the neutron star's is used. In both cases,
photon trajectories are started from a spherical surface with a radius equal to the
neutron star's equatorial radius.

Another type of approximation retains the oblate shape of the rotating neutron star
and embeds the oblate surface in either a Schwarzschild or Kerr spacetime metric. 
We will use the name Oblate Schwarzschild (OS) to denote the use of a Schwarzschild metric
and an oblate surface, and Oblate Kerr (OK) for the use of the Kerr metric and
an oblate surface.

\section{Comparison of Methods}
\label{s:results}

We now turn to the question of how well the various approximations model the
light curves produced by rapidly-rotating neutron stars. We examine three 
aspects of this question. First, we address the effect of the shape of 
the star's surface on the resulting light curve.
Second, we address how the effect of the choice of metric 
affects the resulting light curve.  Third, we examine the errors that
result when the commonly adopted S+D approximation is used to extract the neutron
star's mass and radius values.

\subsection{Oblate Shape of the Star}

The most important factor affecting the light curve of a very rapidly-rotating 
neutron star is the assumed shape of the star's surface. The reason for this can 
be seen in Figure~1, which shows emission from a point at an angle $\theta$ from
the star's spin axis. In Figure~1, the true shape of the star (an oblate spheroid) 
is shown, along with a spherical surface that intersects the point of emission.
In Figure~1, four regions I, II, III, and IV are shown. If  the star's shape is
either spherical or oblate, photons can always be emitted into regions II and III. 
However, in the case of a spherical star, region I is forbidden, and in the case
of an oblate star, region IV is forbidden. If the star is very oblate, the differences
in allowed initial directions can have a very dramatic effect on the resulting
light curve.

\begin{figure}
\begin{center}
\includegraphics[width=3.5in]{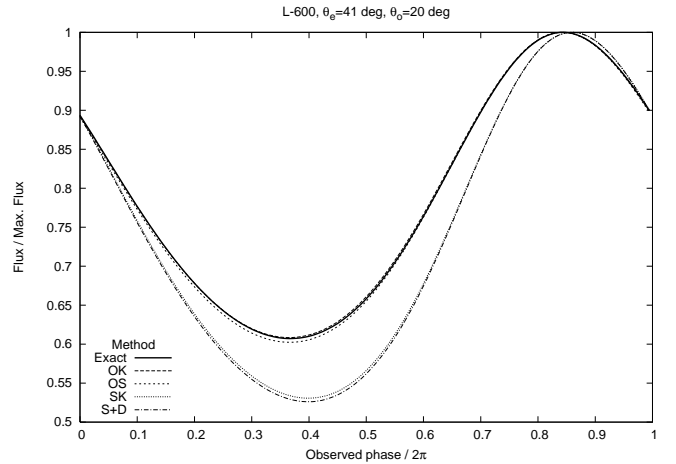}
\end{center}
\caption[]{Light curves for a 600 Hz, 1.4 $M_\odot$ neutron star with EOS L, emission from
an angle of $41^\circ$ from the North pole and an observer at an inclination angle of $20^\circ$
from the North pole. The bold solid curve is the light curve that results if the exact shape 
and gravitational field of the star is used. The four approximation schemes discussed in the
text are shown in various types of dashed lines. Note that the OS and OK curves are very close
to the exact curve and may be difficult to distinguish from the exact case.}
\label{fig:41-20}
\end{figure}

In Figure~\ref{fig:41-20} we show an example for the most oblate star, corresponding to EOS L
with a mass of $1.4 M_\odot$ and a spin frequency of 600 Hz. In Figure~\ref{fig:41-20} we show
the light curves which result if the emission takes place at an angle of $41^\circ$
from the spin axis, and the observer's inclination angle is $20^\circ$ from
the spin axis. The curve labeled ``Exact'' was computed using the 
exact numerical metric and the correct, oblate shape of the star. The curves 
labeled ``OS'' and ``OK'' were calculated by embedding the exact oblate shape
in either the Schwarzschild or Kerr metrics respectively. These three curves are
very close to each other and are difficult to distinguish. The curves labeled
``S+D'' and ``SK'' are computed by assuming a spherical shape for the star
and evolving the photons in either a Schwarzschild or Kerr spacetime. The S+D and 
SK curves are very close to each other, and are very different in shape from the 
curves computed using the oblate surface. The curves computed in the spherical
approximations are more modulated and have a higher harmonic content than the 
curves computed using either the oblate or exact methods. This is purely a geometric effect.
Consider the situation when the spot is on the opposite side of the star from the observer.
In the case of a spherical star the light must be emitted close to tangent to the surface
in order to get to the observer. In the case of an oblate star, the same initial light ray
is emitted closer to the normal to the surface. The solid angle subtended by the spot
is roughly proportional to $\cos \alpha_e$. 
Since the brightness of the spot at any time is roughly approximate to the solid angle subtended by
the spot, the spot's brightness when it is on the far side of the star is larger in the oblate
case than in the spherical case. This leads to less modulation in the case of an oblate star 
compared to a spherical star.

\begin{figure}
\begin{center}
\includegraphics[width=3.5in]{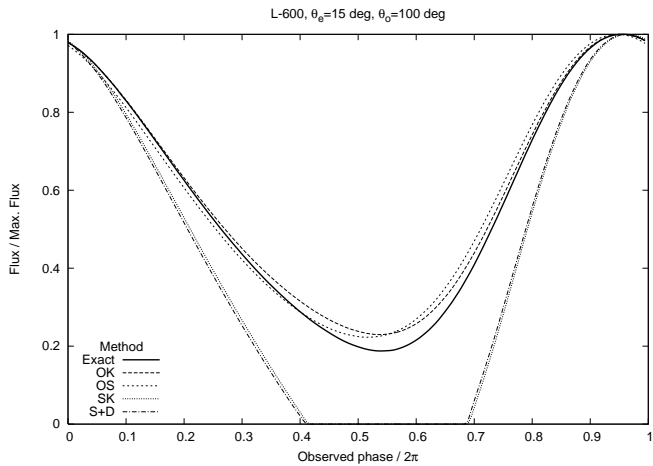}
\end{center}
\caption[]{Light curves for a 600 Hz, 1.4 $M_\odot$ neutron star with EOS L, emission from
an angle of $15^\circ$ from the North pole and an observer at an inclination angle of $100^\circ$
from the North pole.}
\label{fig:15-100}
\end{figure}

In Figure~\ref{fig:15-100} we show a similar set of curves for the same star shown in Figure~\ref{fig:41-20},
but for emission from an angle of $15^\circ$ from the spin pole and an
observer inclination of $100^\circ$. The differences between the curves for
spherical and oblate stellar surfaces are even more extreme than in Figure~\ref{fig:41-20}. 
In Figure~\ref{fig:15-100}, the light curves assuming a spherical surface show eclipses, while
the oblate light curves do not have eclipses. In the spherical cases, the eclipses
occur when light from the back of the star would need to be emitted into region I
(see Figure~\ref{fig:ics}) in order to go over the north end of the star to reach the observer.
Since region I is forbidden when the emission is from a spherical surface, eclipses
occur. In the case of an oblate star, region I is allowed and light reaches the 
observer and no eclipse occurs.

\begin{figure}
\begin{center}
\includegraphics[width=3.5in]{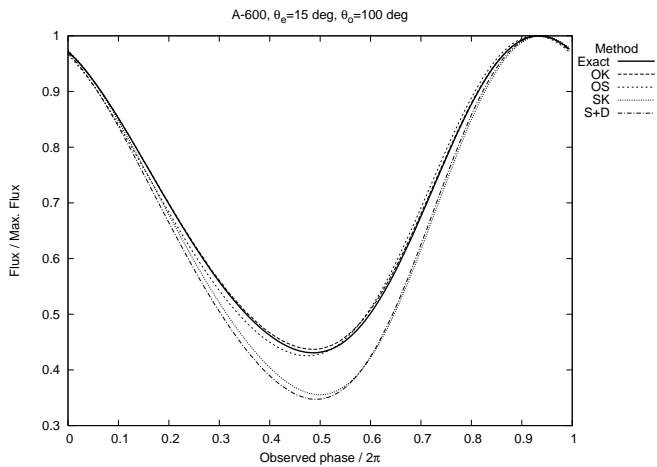}
\end{center}
\caption[]{Light curves for a 600 Hz, 1.4 $M_\odot$ neutron star with EOS A, emission from
an angle of $15^\circ$ from the North pole and an observer at an inclination angle of $100^\circ$
from the North pole.}
\label{fig:A15-100}
\end{figure}

Neutron stars constructed with EOS L are very large, so they are strongly affected by 
rotation and are very oblate. We have chosen to focus on EOS L in order to 
illustrate the largest changes to the waveforms caused by rotation. In contrast, a softer
EOS, such as EOS A produces a smaller star that is less oblate. In Figure~\ref{fig:A15-100}
we show the light curves for the same emission geometry as in Figure~\ref{fig:15-100} but
for EOS A. Note that in the case of the smaller EOS A star, there are no eclipses when the 
spherical surface approximation is made. The shapes of the light curves constructed with
elliptical and spherical models are similar to each other, but the curves constructed with
the spherical approximation are more modulated (for the same reasons as given for the 
case shown in Figure~\ref{fig:41-20}. The lack of eclipses for the spherical star is due to
the fact that the EOS A star is much more compact than the EOS L star (with the same mass).
A more compact star has a stronger gravitational field and the effect of light bending is much
stronger than for a large star, so that it is easier for light to travel from the back side
of the star to the observer. For the remaining figures we have chosen to focus on the larger EOS~L
stars in order to show the largest possible effects due to oblateness, but it should be 
remembered that if the EOS is softer, then the differences between the waveforms computed
using the oblate and spherical models will be smaller than shown.

\begin{figure}
\begin{center}
\includegraphics[width=3.5in]{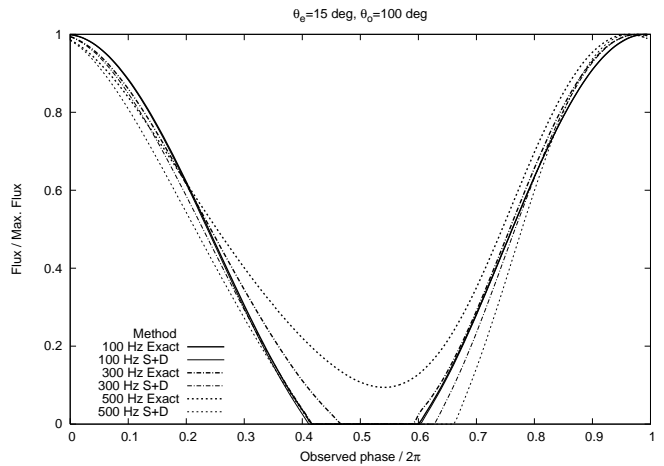}
\end{center}
\caption[]{Light curves for 1.4 $M_\odot$ neutron stars with EOS L, emission from
an angle of $15^\circ$ from the North pole and an observer at an inclination angle of $100^\circ$
from the North pole.  Light curves for spin rates of 100, 300 and 500 Hz are shown. Exact light 
curves are illustrated with bold curves. For comparison the approximate Schwarzschild + Doppler (S+D)
light curves are also shown. The S+D and exact light curves for the 100 Hz star overlap, so the S+D curve
can't be distinguished from the exact in this figure.}
\label{fig:spins}
\end{figure}

In order to illustrate the effect of rotation rate on the light curves, we plot in Figure~\ref{fig:spins}
the light curves for emission from an angle of $15^\circ$ and detection at $100^\circ$ for
stars rotating at frequencies 100, 300 and 500 Hz. For the case of the 100 Hz star (solid lines)
the exact (bold) and approximate S+D (light) curves are almost identical and are difficult to distinguish
from each other in Figure~\ref{fig:spins}. Since 100 Hz is a relatively slow rotation rate and oblateness
is small, the S+D light curve is an excellent approximation to the exact light curve.  The light curves
for 300 Hz stars (dot-dashed lines) show a significant difference between the exact (bold) and
S+D (light) curves. In both cases (for 300 Hz) the spot is eclipsed, but for the exact light curve the 
eclipse lasts for about 1/10 of the spin period, while in the S+D approximation the eclipse lasts
for about 2/10 of the spin period.  In the case of 500 Hz (dotted lines) the exact light curve
(bold) has no eclipse while the S+D approximation (light) does, as in the case of 600 Hz.

\begin{figure}
\begin{center}
\includegraphics[width=3.5in]{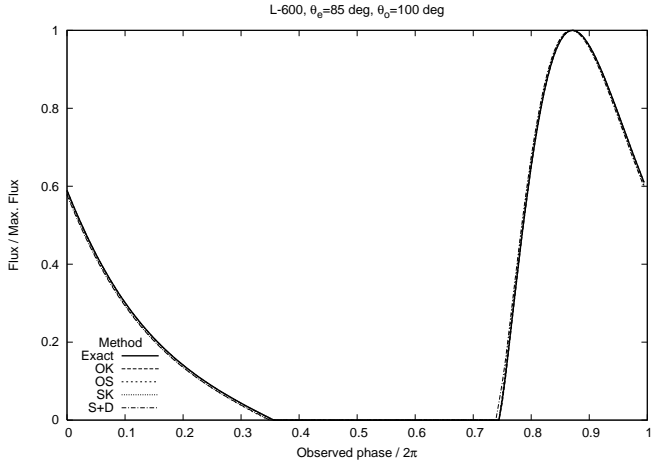}
\end{center}
\caption[]{Light curves for a 600 Hz, 1.4 $M_\odot$ neutron star with EOS L, emission from
an angle of $85^\circ$ from the North pole and an observer at an inclination angle of $100^\circ$
from the North pole.}
\label{fig:85-100}
\end{figure}

When emission and detection takes place near the star's equatorial plane the
differences in the light curves resulting from the 
different calculation methods are not as important, as shown in Figure~\ref{fig:85-100}.
In Figure~\ref{fig:85-100} emission is from a point $85^\circ$ from the spin pole and
the observer has an inclination angle of $100^\circ$. While the fraction
of the time that the signal is ``on'' is different for the oblate and 
spherical cases, the difference is not very important. This lack of difference is
due to the fact that near the equator, there is not much difference between 
the normals to an oblate spheroid or a sphere. This agrees with the earlier
results in CLM for the case of exact equatorial orbits. 

\begin{figure}
\begin{center}
\includegraphics[width=3.5in]{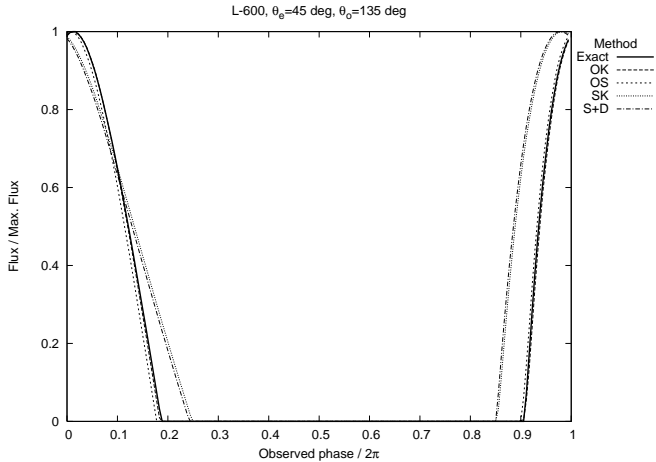}
\end{center}
\caption[]{Light curves for a 600 Hz, 1.4 $M_\odot$ neutron star with EOS L, emission from
an angle of $45^\circ$ from the North pole and an observer at an inclination angle of $135^\circ$
from the North pole.}
\label{fig:45-135}
\end{figure}

When the observer and the emission region are on opposite sides of the star, as in the
case of Figure~\ref{fig:45-135}), the effect of oblateness is to increase the duration
of eclipses compared to an equivalent light curve computed with a spherical surface. The
reason for this can be understood by considering the case of a spherical star. For the
spherical star, the eclipse begins at the moment that the light is emitted tangent to
the surface, into region IV of Figure~\ref{fig:ics}. But the oblate star is not allowed to 
emit into region IV, so this photon never reaches the observer. This analysis holds for
all photons that would have to be emitted into region IV, with the result that the eclipse 
lasts for a longer time in the case of the oblate star. In general, this leads to a
higher asymmetry and harmonic content for light curves for the oblate star (compared
to the spherical star) if the angles of emission and detection are in opposite hemispheres.

Since neither the oblateness of the star nor the emission and detection angles
are typically known for any observation, the use of one of the spherical approximations
for a rapidly-rotating neutron star can lead to incorrect fits. 

\subsection{Metric Approximations}

Given an initial photon direction, the 
most important factor of affecting the propagation of a photon is the strength
of the gravitational potential at the location of emission. To leading order in
rotation, the gravitational potential depends only on the ratio of $M/R$. Approximations
using the Schwarzschild metric to model the neutron star's gravitational field assume that
the value of $M/R$ predicts the deflection and time delay of the photons at a level 
sufficient for light-curve fitting.

The largest rotational correction to the star's metric is the frame-dragging term 
$\omega$ which is of order $\Omega_\star \times M/R$. The use of the Kerr metric
matches the neutron star's metric up to this first order rotational term, but 
does not attempt to match the exact metric at quadratic order (or higher) in
the angular velocity. 

In Figures \ref{fig:41-20}, \ref{fig:15-100} and \ref{fig:85-100}
 we showed the light curves that result when the Schwarzschild 
and Kerr approximations are used. The differences between the choice of metric
are smaller than the differences caused by the shape of the initial emission surface
in the cases illustrated in Figures \ref{fig:41-20} and \ref{fig:15-100}. In the case of 
Figure~\ref{fig:85-100}, all of the
curves are so close that the approximation scheme is unimportant. The uncertainties
in the shape and size of the emission area, along with the uncertainty in the
anisotropy of the emission will produce uncertainties in the shapes of the light
curves which are much larger than the differences in light curves caused by
the choice of metric. However, the errors caused by using the wrong shape of
the surface are competitive with the differences caused by uncertainties in
the emission shape, size and anisotropy. 

As a result, we find no compelling reason to make use of the Kerr metric over the
Schwarzschild metric. An approximation scheme such as the OS results in light
curves that approximate the exact curves about as well as the OK approximation.

\subsection{Errors in Fitted Masses and Radii for the S+D Approximation}

The main motivation for calculating light curves is to extract the neutron star's
mass and radius. At present the only methods for fitting the  mass and radius from light curves
approximate the star's surface as a sphere. We now 
give results of a preliminary investigation of
the errors in fitted mass and
radius resulting from the use of a spherical approximation. First, light curves are created
using the exact (numerical) description of the neutron star for a number
of different stellar models, emission latitudes and observer inclination angles. We then used
a S+D fitting program to deduce the mass and radius of the stars as well as the emission-observer
geometry given the pulse shape and the star's spin frequency. 
In order to isolate the effects of oblateness on the errors, we only used infinitesimal
spot sizes and isotropic emission in both the exact curve generating program and in the S+D 
fitting program. 
This method provides a first clue to the effects of oblateness, but 
in order to model real data a fitting program should be allowed to explore a larger
parameter space including finite spot size, complicated spot patterns and anisotropic 
emission. 

The results of the fits are shown in Table 2. In all cases the true value of the star's mass is $1.4 M_\odot$. 
In this table, the term $R(\theta_e)$ is displayed. When the S+D fitting program fits values, the 
light bending and Doppler boosts depend on the value of the radius at latitude of the spot, not
at the equator. This means that when the S+D fitting program fits the star's radius, it is actually
predicting $R(\theta_e)$ the radial distance from the centre of the star to the fitted location
of the spot. In the column labeled ``$R(\theta_e)$ true'', this is the distance from the centre
of the star to the actual location of the spot. Similarly, the true and fitted values of the
gravitational potential at the location of emission $M/R(\theta_e)$ are shown in the table. 

The S+D fitting is done by selecting a  value of $M/R$ at the location of emission. This fixes the
bending angles and times of arrival in the Schwarzschild metric. Different values of $R$ and the
emission and observer angles are tried, which fix the Doppler boosts, since the star's spin
frequency is assumed to be known. A light curve is calculated and compared to the light curve
computed by the exact method, and  the $\chi^2$ value is calculated
assuming constant error per data point of 0.01, 
with the data normalized so that the peak of the light curve equals 1. 

In the S+D light curves 180 phase bins are used. The mean counts per bin 
ranges from .15 to .7, with peak normalized to 1.0.
The error of 0.01 per phase bin is, using Poisson errors, equivalent to a
220 to 500 counts per bin or a total of 40,000 to 90,000 counts. 
These errors are comparable to those that can be obtained on ms pulsars
using the Rossi X-ray Timing Explorer.  For example, one of the best statistics is
for the light curves shown in  Papitto et al (2005).
They show pulse shapes (mean normalized to 1.0) of SAX J1808.4-3658 
with 64 phase bins and errors per phase bin of 0.005, which implies
40,000 counts per bin for a total of 2.6$\times10^6$ counts. The number of degrees of freedom 
in our fits is 180-NP with NP=6 the number of fit parameters.
Since the S+D model light curve was fitted to exact light curve without added Poisson errors, the
expected $\chi^2$ for a perfect fit is 0 rather than 180-NP. Given the 
assumed errors of 0.01 per phase bin, $\chi^2$ values of 1, 2.71 and 4 roughly correspond to
inconsistency of model and input light curves at the 1$\sigma$, 90$\%$, and 2$\sigma$ levels.   

After the minimum $\chi^2$ value ($\chi^2_{min}$) is found, the best values of the star's radius at the
point of emission and the angles $\theta_e$ and $\theta_o$ given the value of $M/R$ is known.
The process is then repeated for a new value of $M/R$. By iterating this for many values
of $M/R$, the best fit model can be found. In Table 2, the best fit parameters are shown
for each light curve, along with the 90$\%$ uncertainty in the fitted value of $M/R$ and 
the value of $\chi^2_{min}$ for the best-fit model. The uncertainty is computed by finding 
the values of $M/R$ which yield $\chi^2$=$\chi^2_{min}$+2.71. 
We allowed all other parameters to be free in the fits to find the range of parameter of interest
which gave  $\chi^2$=$\chi^2_{min}$+2.71. Thus there have been no assumptions about correlations between 
parameters, and we have found the correct $90\%$ ranges.
In some cases, the minimum in 
$\chi^2$ is so shallow that no meaningful error bars can be computed. We denote these
cases with an asterisk in the uncertainty column. The cases marked with an asterisk are
degenerate, in that almost any values of $M/R$ provides an acceptable fit to the data. 
In some other cases (see emission angle of 45\textdegree, observation angle of 135\textdegree and EOS L) 
the error bars allowed an upper limit on $M/R$ to be made. In these cases the
uncertainty column contains the symbol ``$<$'' and a number which is the upper limit on $M/R$. 
Computing $90\%$ uncertainties for the other parameters than $M/R$ 
is beyond the scope of this paper. One might expect that
in cases where the fits are not degenerate the errors should be of roughly similar fractional 
amount as for $M/R$, however more work is required to confirm this. 

The fit value of $M/R$ often agree closely (within 5\%) with the true $M/R$ values.
Eleven (8 degenerate, 3 non-degenerate) of the fits had M/R fit different 
than $M/R$ true by 10$\%$ to 50$\%$, another 3 had differences between 5 and 10$\%$. If we
discount the degenerate fits, that leaves 6 out of 30 cases with errors between
5$\%$ and 21$\%$. Thus the S+D approximation does have a small but significant (20$\%$) fraction of 
cases with significant errors in $M/R$. 
For the remaining parameters ($M$, $R$ and the angles $\theta_e$, $\theta_o$) the differences between
the true and fit values are larger than for the ratio $M/R$, with usually larger differences
for the angles than for $M$ or $R$. 
For example the fit values of $M$ are different than the true values by  
by 10$\%$ to 24$\%$ in 10 cases (out of 30 nondegenerate fits), and
11 cases had differences between 5 and 10$\%$.
For $\theta_e$, fit values are different than the true values by  
by 10$\%$ to 98$\%$ in 16 cases (out of 30 nondegenerate fits), and
6 cases had differences between 5 and 10$\%$.
Generally, we find that the errors in using the S+D approximation can be large
in some cases and small in some others. However, one must be careful in
interpreting the results of this work: for real data, as previously
mentioned, one would have to generalize
the study to include a finite spot size and more realistic emissivity.

Table 2 illustrates a number of types of fitting problems that occur. For the slowest rotating
stars with spin frequencies $\nu_{star}\le 200$ Hz, the S+D fitting program extracted a value 
of $M/R$ very close to the actual value, as would be expected for slowly rotating stars. The
individual fitted values of mass and radius have some error, but don't deviate by more than
6\% from the true values. In order to fit the radius of the star, the equatorial velocity is fit from
the asymmetry in the light curve due to Doppler boosting. At slow rotation rates the effects
of Doppler boosting is small and the exact light curves have fairly symmetric rise and fall times.
The result is that due to this symmetry a range of radii will fit the data equally well, leading
to a high probability that some error will be introduced into the fitted value of radius. 
As the spin rate is increased, the quality of the fits (as measured by $\chi^2$) decreases,
but the resulting fits for $M/R$ are (within the error) of the correct values of $M/R$ when
the spin frequency is $\le 500$ Hz.  However,
the fits for the masses (or radius) are quite poor in many of these cases. 

In the case of emission from 41\textdegree and observation at 20\textdegree all fits at 
all spin frequencies were degenerate (although we only chose to show a few of these
in Table 2). This can be understood by looking at the light
curve (for the case of 600 Hz) shown in Figure~3. The true light curve for this angle 
combination is very close to a sine wave at all spin frequencies. Since almost any star 
will have a number of emission/observation angle pairs that produce a sine wave
signal, a wide range of model stars provide equally good fits to the data. As others
(eg. \citet{Bhat05}) have remarked, any data which does not contain any harmonics will not 
allow meaningful fits.

Some of the worst errors in extracting mass and radius 
occur for the largest stars constructed with EOS L, one of the stiffest EOS in the 
literature. Since the EOS L stars are very large, they are most affected by rotation and 
are highly oblate. If one is interested in ruling out very stiff equations of state, using
data from rapidly-rotating stars, it is necessary to take the oblateness of these stars into
account.

\section{Conclusions}
\label{s:conclusions}

We have investigated the effect of rapid rotation on the light curves produced by a small spot on
a neutron star's surface. Our calculations are exact, in that we numerically compute the spacetime metric
for the rotating star, include the correct shape of the star and solve 
for the geodesics in the numerical spacetime. We find that the most important 
effect on the light curve arising from rotation comes from the oblate shape of the star's surface.

All past work involving fits of X-ray data have
assumed that the surface of the star is spherical. Our present 
computations show that the assumption of a spherical surface leads to large deviations from the correct
light curve for spin frequencies larger than 300 Hz. However, our test runs show that the assumption
of a spherical surface can still fit the correct value of $M/R$ at spin frequencies $\le 500$ Hz, 
even if the individually fitted values of mass and radius are incorrect.

In our light curve calculations, we find that if the star's surface is assumed to be a sphere,
the light curves resulting from the use of the Kerr metric are almost the same as those
resulting from the use of the Schwarzschild metric. The difference in these light curves is smaller
than the size of the error bars in any present data, so it seems that there is no
significant advantage in the use of the SK approximation over the S+D approximation.

We calculated light curves in an approximation where an oblate surface is embedded in
either the Schwarzschild (OS) or Kerr (OK) spacetime. In both cases, the resulting
light curves are very close to the exact light curves computed using the numerical
spacetime of the rotating neutron star. This suggests that an approximation similar
to the Schwarzschild + Doppler approximation that includes an oblate surface could
be a useful way to fit data. The construction of a fitting program that extracts the
mass and radius of a star from a light curve using an oblate surface is non-trivial.
In future work, we plan to develop a code that will allow an oblate shape to be 
incorporated into a S+D fitting program that includes finite spot sizes
and anisotropic emission. We anticipate that such a program will
allow the accurate extraction of the masses and radii of very rapidly-rotating 
neutron stars.

\acknowledgments
We thank the referee for comments that helped us clarify the presentation
of our results. This research was supported by grants from NSERC.

\begin{deluxetable}{cllllll}
\tablecaption{Neutron Star Parameters for 1.4 $M_\odot$ Stars}
\tablewidth{0pt}
\tablehead{
\colhead{EOS} & \colhead{$\nu_*$} & \colhead{$R${\tablenotemark{a}}} & 
\colhead{$GM/Rc^2$} & \colhead{$Jc/M^2G$} & \colhead{$v/c${\tablenotemark{b}}} 
& \colhead{$\nu_B$\tablenotemark{c}}\\
\colhead{}    & \colhead{Hz} &      \colhead{km}  &
\colhead{}          & \colhead{}          & \colhead{}      & \colhead{Hz}
}
\startdata
A &   0 & 9.57 & 0.216 & 0    & 0    & 1390 \\
  & 100 & 9.57 & 0.216 & 0.04 & 0.03 &       \\
  & 200 & 9.59 & 0.216 & 0.07 & 0.05 & \\
  & 300 & 9.62 & 0.215 & 0.11 & 0.08 & \\
  & 400 & 9.66 & 0.214 & 0.15 & 0.11 & \\
  & 500 & 9.71 & 0.213 & 0.19 & 0.13 & \\
  & 600 & 9.78 & 0.211 & 0.22 & 0.16 & \\
L &   0 & 14.8 & 0.139 & 0    & 0    & 740 \\
  & 100 & 14.9 & 0.139 & 0.08 & 0.04 & \\
  & 200 & 14.9 & 0.138 & 0.15 & 0.07 & \\
  & 300 & 15.1 & 0.137 & 0.23 & 0.11 & \\
  & 400 & 15.4 & 0.135 & 0.32 & 0.15 & \\
  & 500 & 15.7 & 0.131 & 0.41 & 0.19 & \\
  & 600 & 16.4 & 0.126 & 0.51 & 0.24 & \\
\enddata
\tablenotetext{a}{Equatorial radius}
\tablenotetext{b}{Equatorial speed measured by a static observer at the star's surface} 
\tablenotetext{c}{Break-up spin frequency for a star with the given mass and EOS}
\label{tab:1}
\end{deluxetable}

\begin{deluxetable}{ll|c|c||c||r|r||r||r||l|l|l||r}
\tabletypesize{\footnotesize}
\tablecaption{Comparison of True and Fitted Parameters of Neutron Stars. All neutron
stars have a mass of $1.4 M_\odot$.}
\tablehead{
\colhead{$\theta_e$}&
\colhead{$\theta_o$}&
\colhead{EOS} & 
\colhead{$\Omega_{\star}$ (Hz)} & 
\colhead{$M/M_{\odot}$} & 
\multicolumn{2}{||c||}{$R(\theta_e)$ (km)} & 
\colhead{$\theta_e$ (deg)} &
\colhead{$\theta_o$ (deg)} & 
\multicolumn{3}{||c||}{$GM/c^2 R(\theta_e)$} & 
\colhead{$\chi^2$} \\
\multicolumn{2}{c}{true}&
\colhead{}&
\colhead{}&
\colhead{fit} &
\colhead{true} & \colhead{fit} &
\colhead{fit} &
\colhead{fit}  &
\colhead{true} & \colhead{fit} & \colhead{unc.} & \colhead{}} 
\startdata
41\textdegree & 100\textdegree &

   A & 100 & 1.48 &9.57 & 10.2 & 80.5 & 139.2 & 0.216 & 0.215 & 0.011 & 0.1 \\ 
&& L &     & 1.32 &14.83& 13.8 & 80.5 & 133.0 & 0.140 & 0.142 & 0.027      & 0.02 \\ 
&& A & 300 & 1.49 &9.58 & 10.0 & 79.8 & 138.0 & 0.216 & 0.220 & 0.005 & 1   \\ 
&& L &     & 1.09 &14.82& 11.1 & 67.0 &  95.6 & 0.140 & 0.145 & 0.024 & 0.3 \\ 
&& A & 400 & 1.45 &9.58 & 9.55 & 80.8 & 134.9 & 0.216 & 0.225 & 0.006 & 2   \\ 
&& L &     & 1.17 &14.80& 11.9 & 58.0 & 96.3  & 0.140 & 0.145 & 0.023 & 0.4 \\ 
&& A & 500 & 1.51 &9.59 & 9.89 & 80.2 & 136.9 & 0.216 & 0.225 & 0.005 & 3   \\ 
&& L &     & 1.29 &14.78& 12.7 & 52.7 & 98.1  & 0.140 & 0.15  & 0.02  & 0.8 \\ 
&& A & 600 & 1.58 &9.60 & 10.2 & 41.9 & 102.2 & 0.215 & 0.230 & 0.007 & 4   \\ 
&& L &     & 1.30 &14.74& 12.0 & 57.9 & 97.5  & 0.140 & 0.160 & 0.015 & 2   \\ 
\\
85\textdegree & 100\textdegree &                
   A & 100 & 1.48 & 9.57  & 10.4 & 87.4 & 110.8 & 0.216 & 0.210  & 0.008   & 1    \\ 
&& L &     & 1.45 & 14.86 & 15.3 & 84.0 & 103.7 & 0.139 & 0.140  & 0.027  & 0.05 \\ 
&& A & 200 & 1.46 & 9.59  & 10.1 & 84.4 & 103.9 & 0.216 & 0.215 & 0.006  & 2    \\ 
&& L &     & 1.43 & 14.95 & 15.6 & 86.7 & 107.6 & 0.138 & 0.135 & 0.029   & 0.4  \\ 
&& A & 300 & 1.48 & 9.62  & 10.2 & 83.1 & 103.5 & 0.215 & 0.215 & 0.025  & 4    \\ 
&& L &     & 1.70 & 15.10 & 17.9 & 80.0 & 123.3 & 0.137 & 0.140  & 0.015  & 0.4  \\ 
&& A & 400 & 1.49 & 9.66  & 10.3 & 77.2 & 99.7  & 0.214 & 0.215 & 0.003  & 7    \\ 
&& L &     & 1.40 & 15.35 & 16.0 & 85.0 & 105.5 & 0.135 & 0.130  & 0.009   & 4    \\ 
&& A & 500 & 1.68 & 9.71  & 11.3 & 67.7 & 111.0 & 0.213 & 0.220  & 0.004  & 5    \\ 
&& L &     & 1.51 & 15.73 & 17.9 & 61.7 & 97.7  & 0.131 & 0.125 & 0.009   & 6    \\ 
&& A & 600 & 1.62 & 9.78  & 11.1 & 72.5 & 113.6 & 0.211 & 0.215 & 0.003 & 0.1  \\ 
&& L &     & 1.40 & 16.35 & 17.3 & 77.8 & 102.6 & 0.127 & 0.120  & 0.007 & 0.2  \\ 
\\
45\textdegree & 135\textdegree &
   A & 100 & 1.51 & 9.57   & 10.4 & 47.4 & 137.8 & 0.216  & 0.216       & 0.005   & 0.1  \\ 
&& L &     & 1.46 & 14.83  & 15.4 & 44.8 & 135.4 & 0.139  & 0.140       & $<0.155$& 0.06 \\ 
&& A & 200 & 1.60 & 9.58   & 10.9 & 45.5 & 138.5 & 0.216  & 0.218       & 0.004   & 1    \\ 
&& L &     & 1.48 & 14.84  & 15.7 & 44.3 & 136.0 & 0.139  & 0.140       & $<0.145$& 0.1  \\ 
&& A & 300 & 1.57 & 9.58   & 10.7 & 45.1 & 137.6 & 0.216  & 0.217       & 0.006   & 0.5  \\ 
&& L &     & 1.45 & 14.85  & 15.9 & 44.0 & 136.0 & 0.139  & 0.135       & $<0.155$& 0.1  \\ 
&& A & 400 & 1.73 & 9.59   & 11.6 & 41.6 & 139.0 & 0.216  & 0.220       & 0.005   & 8    \\ 
&& L &     & 1.44 & 14.87  & 16.3 & 43.0 & 136.4 & 0.139  & 0.130       & $<0.145$& 0.1  \\ 
&& A & 500 & 1.61 & 9.61   & 11.0 & 44.5 & 138.6 & 0.215  & 0.216       & 0.005   & 1    \\ 
&& L &     & 1.22 & 14.81  & 18.0 & 32.0 & 123.4 & 0.139  & 0.100       & $<0.135$& 0.8  \\ 
&& A & 600 & 1.63 & 9.63   & 11.2 & 42.5 & 137.7 & 0.215  & 0.215       & 0.005   & 1    \\ 
&& L &     & 1.34 & 14.90  & 18.0 & 40.0 & 137.5 & 0.139  & 0.110       & $<0.145$& 0.6  \\ 
\\
41\textdegree & 20\textdegree &
   A & 100 & 1.41 & 9.57  & 9.23  & 29.8 & 28.9 & 0.216 & 0.225 & * & 0.005 \\ 
&& A & 500 & 1.40 & 9.59  & 10.4  & 20.6 & 35.3 & 0.216 & 0.2   & * & 0.05  \\ 
&& L &     & 1.99 & 14.78 & 16.8  & 33.2 & 21.5 & 0.140 & 0.175 & * & 0.03  \\ 
&& A & 600 & 1.40 & 9.60  & 10.9  & 20.2 & 34.0 & 0.215 & 0.19  & * & 0.07  \\ 
&& L &     & 2.53 & 14.74 & 17.8  & 28.8 & 23.2 & 0.140 & 0.21  & * & 0.06  \\ 
\\
15\textdegree & 100\textdegree &
   A & 100 & 1.08  & 9.57  & 6.35  & 30.1 & 80.9 &0.216 & 0.25 & * & 4   \\ 
&& A & 500 & 0.59 & 9.51  & 8.76  & 54.7 & 21.8 &0.217 & 0.1  & * & 1   \\ 
&& L &     & 0.85 & 14.25 & 8.41  & 30.2 & 78.1 &0.145 & 0.15 & * & 0.8 \\ 
&& A & 600 & 0.68 & 9.49  & 9.10  & 56.9 & 20.3 &0.218 & 0.11 & * & 0.8 \\ 
&& L &     & 0.92 & 13.98 & 7.98  & 34.1 & 69.5 &0.148 & 0.17 & * & 2   \\ 
\enddata
\label{tab:2}
\end{deluxetable}


\appendix

\section{Relation Between Times-of-Arrival and Deflection Angles}
\label{s:appendix}

In order to compute the coordinate time elapsed during the photon's flight from the star to
the observer, equation (\ref{teq}) must be integrated from the initial to final values of
the affine parameter $\lambda$. This is complicated by the fact that the metric potentials
$\gamma, \rho$ and $\omega$ appearing in equation (\ref{teq}) depend on the values of
the coordinates $\rb(\lambda)$ and $\theta(\lambda)$ describing the location of the photon at 
every point along the geodesic. 

It is useful to introduce a flat, spacelike two-dimensional space perpendicular to the 
full $3+1$ spacetime's Killing vectors. As the photon moves through the real
spacetime it also traces out a curve in the two-space. The arc-length along the curve in
the two-space is denoted $\zeta$ and is given by the usual flat space formula
\be
d\zeta^2 = d\rb^2 + \rb^2 d\theta^2.
\ee
Since the arc-length depends on the value of the affine parameter $\lambda$, 
equation (\ref{eq:nullconst}) can be used to write an equation for $\dot{\zeta}\equiv d\zeta/d\lambda$,
\be
\dot{\zeta}^2 = {\mathcal A}(\rb(\zeta),\theta(\zeta)).
\ee

The elapsed coordinate time can now be written as the integral
\begin{eqnarray}
T(\theta_i,\theta_f,b)  &=& 
\int \frac{dt}{d\zeta} d\zeta = \int \frac{\dot{t}}{\dot{\zeta}} d\zeta 
= \int {{\mathcal A}^{-1/2}}{e^{-(\gamma+\rho)}(1-\omega b)} d\zeta.
\label{a:teq}
\end{eqnarray}
Similarly, the azimuthal deflection can be written as
\be
\psi(\theta_i,\theta_f,b) 
= \int {{\mathcal A}^{-1/2}}\left({\omega e^{-(\gamma+\rho)}(1-\omega b)+e^{\rho-\gamma}\frac{b}{\rb^2\sin^2\theta}} \right) d\zeta.
\label{a:phieq} 
\ee
Neither equation (\ref{a:teq}) nor (\ref{a:phieq}) is in a form useful for directly calculating
either the elapsed time or the azimuthal deflection. However, note that both of these equations
are of the form of exact line integrals. If the value of the impact parameter is kept fixed,
the values of the two integrals depend only on the endpoints. If the endpoints of the 
geodesic are kept fixed, the changes in the arrival time and in the azimuthal deflection
angles can easily be calculated for small changes in the impact parameter. 
Making use of equations (\ref{Beq}) and (\ref{eq:nullconst}), the derivative of the
coordinate arrival time with respect to the impact parameter is found to have the value
\begin{eqnarray}
\frac{dT}{db} &=& 
\int {\mathcal A}^{-3/2} e^{-2\alpha-2\gamma} \frac{b}{r^2 \sin^2\theta} d\zeta.
\label{dTdb}
\end{eqnarray}
Similarly, we find 
\begin{eqnarray}
\frac{d\psi}{db} &=& \int {\mathcal A}^{-3/2} e^{-2\alpha -2\gamma}  \frac{1}{r^2 \sin^2\theta} d\zeta.
\label{dpsidb}
\end{eqnarray}
Comparing equations (\ref{dTdb}) and (\ref{dpsidb}) we are led to the identity
\be
\frac{dT}{db}(\theta_i,\theta_f,b) = b \frac{d\psi}{db}(\theta_i,\theta_f,b).
\label{identity}
\ee
As a result, if two photons are emitted from the same latitude $\theta_i$ on the star
and are detected by the same observer, the difference in arrival times is related 
to the difference in the azimuthal deflection by the simple formula given by
equation (\ref{identity}).

As a quick check, we note that in the limit of zero rotation the metric potentials
are given by equations (3 - 5) of Cadeau et al (2005), and the integrals
for the time delay and azimuthal deflection reduce to simple integrals over the
radial coordinate. It is easy to explicitly check that the identity (\ref{identity}) holds
in the case of the Schwarzschild metric, as it must since our proof holds for 
any axisymmetric stationary spacetime.


\end{document}